\documentclass[prl,twocolumn,superscriptaddress,showpacs]{revtex4-1}
\usepackage{graphicx}
\usepackage{color}
\usepackage{physics}

\newcommand*{\I}{\imath}

\newcommand*{\D}{\mathrm{d}}

\newcommand {\IM}{\operatorname{Im}}
\newcommand{\e}{\varepsilon}
\newcommand{\W}{\Omega}

\begin{document}
\title{Locking and regularisation of chimeras by periodic forcing}
\author{Maxim I. Bolotov}
\affiliation{Department of Control Theory, Nizhny Novgorod State University, Gagarin Av.\ 23, 606950, Nizhny Novgorod, Russia}
\author{Lev A. Smirnov}
\affiliation{Department of Control Theory, Nizhny Novgorod State University, Gagarin Av.\ 23, 606950, Nizhny Novgorod, Russia}
\affiliation{Institute of Applied Physics of the Russian Academy of Sciences, Ul'yanov str.\ 46, 603950, Nizhny Novgorod, Russia}
\author{Grigory V. Osipov}
\affiliation{Department of Control Theory, Nizhny Novgorod State University, Gagarin Av.\ 23, 606950, Nizhny Novgorod, Russia}
\author{Arkady Pikovsky}
\affiliation{Institute of Physics and Astronomy, University of Potsdam, Karl-Liebknecht-Stra\ss{}e 24-25, 14476 Potsdam, Germany}
\affiliation{Department of Control Theory, Nizhny Novgorod State University, Gagarin Av.\ 23, 606950, Nizhny Novgorod, Russia}
\affiliation{Laboratory of Topological Methods in Dynamics, HSE University, Russia} 
\begin{abstract}
We study how a chimera state in a one-dimensional medium of non-locally coupled
oscillators responses to a periodic external force. On a macroscopic level, where chimera
can be considered as an oscillating object, forcing leads to entrainment of the chimera's frequency inside an Arnold tongue.
On a mesoscopic level, where chimera can be viewed as a inhomogeneous, stationary or nonstationary pattern,
strong forcing can lead to reguralization of an unstationary chimera. On a microscopic level of the dynamics of individual oscillators,
forcing outside of the Arnold tongue leads to a multi-plateu state with nontrivial locking properties. 
\end{abstract}

\maketitle
Chimeras in non-locally coupled oscillator populations
are spectacular patterns combining synchronous and asynchronous patches. 
Since the first observation by 
Kuramoto and Battogtokh (KB)~\cite{Kuramoto-Battogtokh-02},
a significant progress has been achieved in 
theoretical and experimental
exploration of chimeras, see recent reviews~\cite{Panaggio-Abrams-15,Omelchenko-18}. 
On a microscopic level, the KB-chimera demonstrates
coexistence of ordered and disordered domains: neighboring units
are either fully synchronized or partially correlated. 
On a mesoscopic level, when one
introduces a coarse-grained order parameter, 
chimera constitutes a nonhomogeneous pattern with
a continuous profile of the complex order parameter,
the latter has modulus one in synchronized domains and is less than
one in disordered regions. 
On a macroscopic level,
chimera can be treated as an inhomogeneous oscillating structure. 

In 
a homogeneous medium, because of the invariance to shifts in space and time,
chimera pattern is expected to be sensitive to a weak inhomogeneity in space and to
a small time-dependent forcing. The former option was explored in 
refs.~\cite{Bick-Martens-15,Ruzzene_etal-19}, where it has been demonstrated 
that a weak inhomogeneity controls chimera's position in space. 
In this
paper we focus on a time-periodic, uniform in space forcing.
The main guiding point is that the chimera as whole is an oscillating
object, and thus, similar to simple self-sustained
oscillators,  can be phase locked or frequency 
entrained. In particular, two coupled chimera can synchronize
in the sense of entrainment of their mean frequencies, while the internal inhomogeneous
structure of order-disorder is preserved~\cite{Andrzejak_etal-18}. 

In this paper we explore synchronization properties of chimera patterns subject to a periodic external
force. We employ the reduction approach of \cite{Smirnov-Osipov-Pikovsky-17}, and formulate the problem
of finding chimera patterns locked to the external field as a problem of finding periodic
orbits in a system of ordinary differential equations (ODEs). At this macroscopic
level, we determine
chimeras with stable and unstable phase shifts to the forcing, and regions of
their existence (``Arnold tongues'' (AT)) on the plane of forcing parameters 
``amplitude -- frequency''. On the mesoscopic level, locked chimeras can be
stationary, breathing, or turbulent, and we characterize these state through distributions
of the oscillators mean frequencies. Outside of the locked region, interesting
microscopic patterns appear, where some subgroups of oscillators
are mutually entrained, and some of them are entrained by the external force (while the chimera as a whole
is not).


As a basic model we use the KB setup~\cite{Kuramoto-Battogtokh-02} and
consider a one-dimensional (1D) oscillatory medium of length $L$, enclosed in a ring and described
by phases $\varphi(x,t)$ which are coupled non-locally. This
medium is additionally subject to a uniform force with amplitude $\varepsilon$ and
frequency{\,}$\Omega$:
\begin{equation}
\!\partial_{t}\varphi\!=\!\IM\!\left(\!
e^{-\I(\varphi+\alpha)}\!
\int_0^L\!\!\!G(x\!-\!\tilde x)e^{\I\varphi(\tilde x,t)}\D \tilde x + 
\e e^{\I(\W t - \varphi)}
\!\right)\!.\!\!\!\!
\label{eq:be}
\end{equation}
Here the kernel $G(x)={\cosh\bigl(|x|-L/2\bigr)}\bigl/{2\sinh\bigl(L/2\bigr)}\bigr.$ is 
a regularized (for periodic boundary condition) version of exponential kernel
used by KB~\cite{Kuramoto-Battogtokh-02,Smirnov-Osipov-Pikovsky-17} 
[this kernel is the inverse of the operator $(\partial_{xx}-1)$
in the periodic domain while the exponential kernel is the inverse 
in the infinite domain, cf. Eq.~\eqref{eq_pdeH} below].

\looseness=-1
The first step in the analysis is introduction of
the coarse-grained order parameter $Z(x,t) \!=\! \langle e^{\I\varphi(x,t)} \rangle_{\text{loc}}$
by using the procedure of averaging over a small vicinity of the point $x$~\cite{Ott-Antonsen-08,Laing-09,Bordyugov-Pikovsky-Rosenblum-10,Smirnov-Osipov-Pikovsky-17}.
Physically, the amplitude of this complex function (which satisfies the inequality $|Z(x,t)| \leq {1}$)
fully describes the level of synchrony of the phases $\varphi(x,t)$ in a smallz neighborhood of the point $x$.
In region where $|Z(x,t)|={1}$, the neighboring elements move synchronously.
When  $|Z(x,t)|<{1}$ the neighboring phase oscillators rotates asynchronously.
The complex order parameter $Z(x,t)$ obeys the Ott-Antonsen
equation~\cite{Ott-Antonsen-08,Laing-09,Bordyugov-Pikovsky-Rosenblum-10,Smirnov-Osipov-Pikovsky-17}
\begin{gather}
\partial_t Z = (e^{-\I \alpha}H - e^{\I \alpha}H^{*}Z^2) / 2,\label{eq_Z}\\
H(x,t) = \varepsilon e^{\I (\Omega t + \alpha)}+\!\int_{0}^{L}\!\!\!G(x-\tilde{x})
Z(\tilde{x},t)\dd\tilde{x}.
\label{eq_IntH}
\end{gather}
Equation \eqref{eq_IntH} can be equivalently written as a partial differential equation (PDE)
\begin{equation}
\partial_{xx}^2 H-H = -Z-\varepsilon e^{\I (\Omega t + \alpha)}.
\label{eq_pdeH}
\end{equation}
Thus, the problem of finding forced chimera states can be formulated as that
of finding nontrivial patterns in the system of 
PDEs~\eqref{eq_Z},\eqref{eq_pdeH}.


We look for patterns uniformly rotating with the frequency of 
forcing: $Z(x,t)=z(x)e^{\I \Omega t}$, $H(x,t)=h(x)e^{\I \Omega t}$.
This yields a system of equations for the spatial profiles
\begin{equation}
e^{\I \alpha}h^{*}z^2+2\I\Omega z-e^{-\I\alpha}h=0,\quad
h''-h=-z-\varepsilon e^{\I \alpha}.
\label{eq_h_z}
\end{equation}
Hereafter, by primes at the functions of the variable $x$ we will denote derivatives with respect to the spatial coordinate $x$.
Expressing from the first algebraic equation $z$ through $h$ (where one of two roots is chosen
according to the local stability condition), we obtain the final
4-th order ODE system
\begin{equation}
h''-h=\displaystyle \I \dfrac{\Omega + \sqrt{\Omega^2 - |h|^2}}{e^{\I \alpha}h^{*}}-\varepsilon e^{\I\alpha}.
\label{eq_h}
\end{equation}
This equation for the autonomous case ($\e=0$)
has been explored in \cite{Smirnov-Osipov-Pikovsky-17} to find free (unforced) chimera patterns. 
The main difference to the free case
is that the forcing term $\e$ breaks the
phase shift invariance $\theta\!\to\!\theta\!+\!\theta_0$, 
where $\theta\!=\!\text{arg}(h)$. The latter invariance
allowed us to reduce Eq.~\eqref{eq_h} to a 3-dimensional system in the absence of forcing,
but now the full 4-th order system is to be\,considered.

Our strategy (described in more details in \footnote{see Supplementary Material (SM)})
is to find, for each pair of values $\e,$ $\W$, a symmetric
periodic solution of~\eqref{eq_h} starting from a state $|h|(0)$, $\theta(0)$,
$|h|'(0)=0$, $\theta'(0)\!=\!0$. The initial value $|h|(0)$ is adjusted to find a periodic
solution, period of which $L$ will depend on the initial phase $\theta(0)$. As
the dependence $L(\theta(0))$ is a $2\pi$-periodic function of $\theta(0)$, in a certain
range of system lengths $L_{\min}<L<L_{\max}$ we have at least two solutions 
for each $L$ (in fact,
in all cases presented below this number is exactly two because we consider only 1:1
locking). We illustrate this in Fig.~1 of SM.

The next step is determining $L_{\min}(\e, \W)$ and $L_{\max}(\e,\W)$ in a range
of values of the forcing frequency $\W$, for fixed $\e$. These curves are also
illustrated in Fig.~2 of SM. To find the borders of the AT, i.e.
the synchronization region, for a fixed length of the medium $L$, we have 
to inverse these dependencies as $\W_{\min}(\e,L)$ and $\W_{\max}(\e,L)$. Then,
the phased locked solutions for a chimera pattern in the system
of length $L$ exist in the AT defined 
as $\W_{\min}(\e,L)<\W<\W_{\max}(\e,L)$. As $\e\to 0$, this tongue
shrinks to the frequency $\W_0(L)$ of the autonomous chimera pattern.

\begin{figure}[t]
\includegraphics[width = \columnwidth]{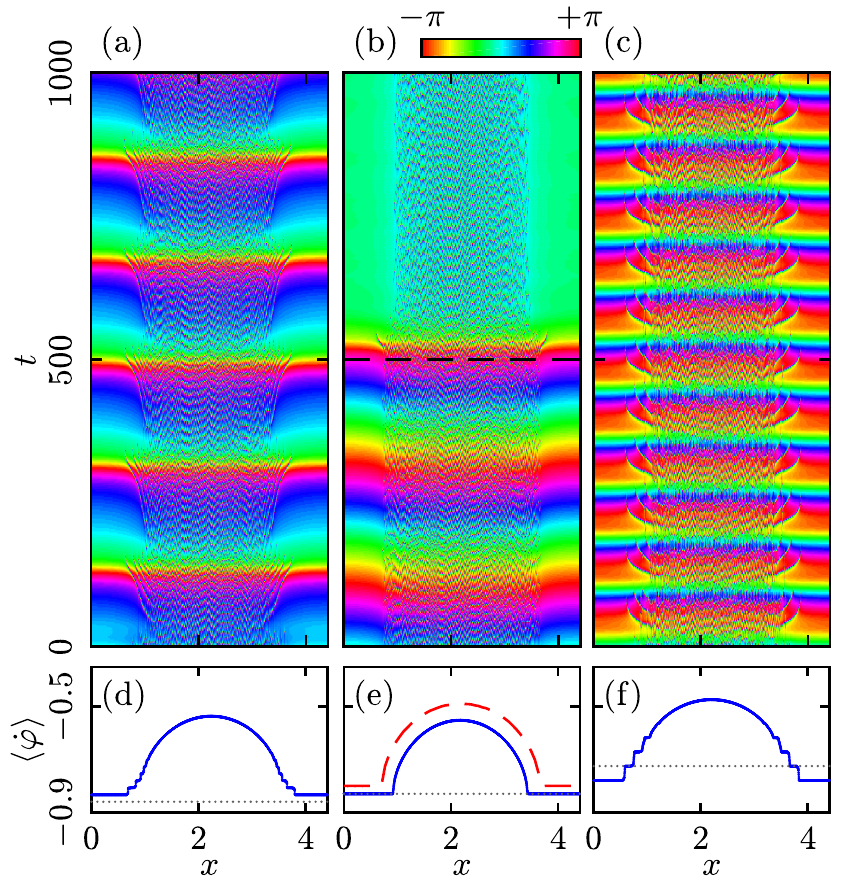}
\caption{Direct numerical simulations of the set of $N=4096$ oscillators performed 
within the phase model~\eqref{eq:be} with the 
parameters $\varepsilon = 0.025$ and  $\Omega = -0.86$ (a,d), 
$\Omega = -0.83$ (b,e), and $\Omega = -0.725$ (c,f).
Panels (a,b,c): spatio-temporal plot of the phases in the reference 
frame rotating with an angular velocity $\Omega$.
Panels (d,e,f): average frequencies of the elements (blue lines)
together with the forcing frequency $\Omega$ (gray doted line).
In each case, the initial state was close to an autonomous single-cluster 
chimera state at the length $L \approx 4.41$ and the corresponding natural 
frequency $\Omega_{0}=-0.8$.
For situations depicted in panels (a) and (c), an external force was 
present during the full simulation time.
In the case shown in panel (b), the force was switched on abruptly at an instant 
of time $t_0=500$ (black dashed straight line in panel (b)).
A red dashed curve in panel (e) shows a profile of average 
frequencies of the oscillators for a free-runing chimera state.}
\label{fig:atst} 
\end{figure}

\begin{figure}[!ht]
\includegraphics[width = 0.8\columnwidth]{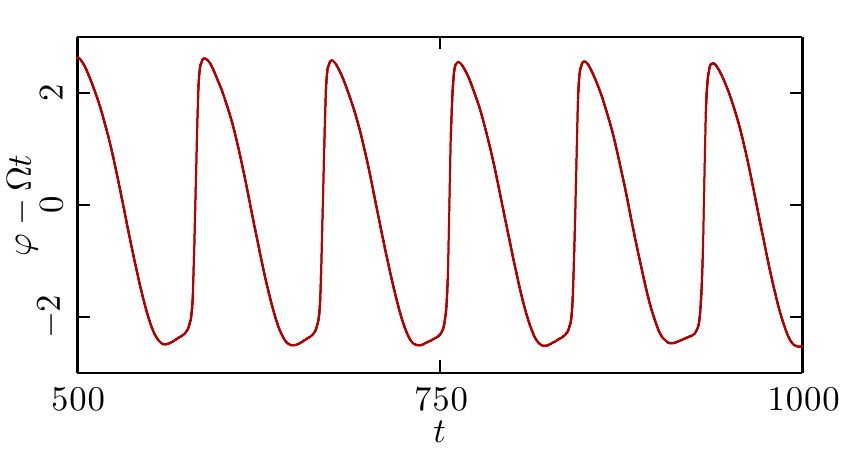}
\caption{The dynamics of the phase of an oscillator from the subgroup of\,Fig.\,\ref{fig:atst}(c), 
entrained by the external force, $x\!=\!0.64$. }
\label{fig:ph} 
\end{figure}

Additionally, we have to check general stability of the found patterns.
To this end one linearizes the full system~\eqref{eq_Z},~\eqref{eq_pdeH}
and solves the eigenvalue problem. Practically, this can be done via
a finite-difference representation described in~\cite{Smirnov-Osipov-Pikovsky-17} and SM, 
allowing for reducing the problem to that of finding eigenvalues of large matrices. The resulting spectrum
of eigenvalues has a continuous branch (essential spectrum) and discrete
eigenvalues, which are responsible for possible instabilities (see detailed
discussion in Refs.~\cite{Omelchenko-13,Xie_etal-14}). 

The stability analysis can be also applied to autonomous chimera patterns~\cite{Smirnov-Osipov-Pikovsky-17},
correspondingly one distinguishes stable and unstable chimeras. The latter solutions
evolve typically into breathing (time-periodic)~\cite{Kemeth_etal-16,Bolotov_etal-17a,Suda-Okuda-18,Bolotov-Smirnov-Osipov-Pikovsky-18} or turbulent 
chimeras~\cite{Bordyugov-Pikovsky-Rosenblum-10,Bolotov_etal-17a}. Below we discuss the effect of 
periodic forcing on these states. We fix
the value of parameter $\alpha=1.457$, so the only parameter of a free chimera  to vary
is the system length $L$; the natural frequency $\Omega_0$  is uniquely determined by $L$.

Let us first consider the effect of external periodic forcing on a stable chimera. 
We exemplify this case with parameters $L\!=\!4.41$, $\W_0\!=\!-0.8$. 
Here, for small forcing amplitudes $\e \!\lesssim\! 0.05$ we obtain a standard AT
on the plane $\e$, $\W$, inside which a locked chimera is stable (see Fig.\,3 of SM).
The dynamics of locking is illustrated in Fig.\,\ref{fig:atst}(b). Here we show the phases
of oscillators in a free-running state until time $t_0\!=\!500$, at which the forcing with frequency
$\Omega \!=\! -0.83$ and amplitude $\e\!=\!0.025$ is applied. The phases are shown in the reference frame rotating with
the external frequency $\Omega$, thus for $t\!<\!t_0$ one observes rotation of the phase in the synchronous domain.
The effect of locking is evident by
inspection of the phase difference between the external force and the coherent domain
for $t\!>\!t_0$: in the locked  state it is constant. This corresponds to the shift of 
the profile
of average frequencies of all the oscillators (Fig.\,\ref{fig:atst}(e)): the whole profile shifts so
that the frequency of the coherent domain becomes exactly 
the external one (depicted by the dotted line).

\looseness=-1
Outside of the locking region (i.e.
for a large mismatch between $\W$ and $\W_0$), one observes unlocked quasiperiodic regimes 
(Fig.~\ref{fig:atst}(a),(c)). 
For $\W<\W_0$, all the oscillators have frequencies larger than $\W$, one can clearly see 
the plateaus at the modulation frequency and its harmonics in the profile of average frequencies
(Fig.~\ref{fig:atst}(d)).
For $\W>\W_0$, the major part of the coherent oscillators have a frequency less than
$\W$, but there exists another plateau exactly at the driving frequency (Fig.~\ref{fig:atst}(f)).
Remarkably, the phases of these oscillators, in the reference frame rotating with $\W$,
are not constants, but experience rather large variations (see Fig.~\ref{fig:ph}) -- 
nevertheless, they are perfectly frequency entrained by the force.
The existence of the plateaus in the frequency profile resembles that for 
breathing chimeras~\cite{Kemeth_etal-16,Bolotov_etal-17a}. In the latter case, however,
an extra modulation frequency appears due to instability of the stationary chimera;
in our case the modulation frequency is due to an imperfect locking to the external field.


\begin{figure}[ht]
\includegraphics[width = \columnwidth]{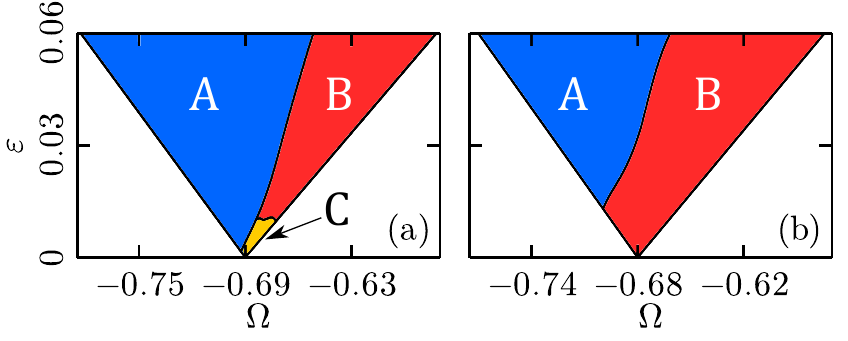}
\caption{Existence domains of locked chimera patterns (AT) 
for (a) $L \approx 6.854$ and (b) $L \approx 7.332$. here unforced chimeras with
natural frequencies (a) $\Omega_{0}=-0.69$ 
and (b) $\Omega_{0}=-0.68$, are unstable.  Inside AT, in region $A$ (blue color)
a locked chimera is stable. 
In regions $B$,\,$C$ (red and yellow colors), all stationary chimeras are unstable,
and the observed state is either turbulent in region $B$ (see Fig.~\ref{fig:atinst}
below) or time-periodic (breathing chimera, region $C$).}
\label{fig:oscd} 
\end{figure}

\begin{figure}[!htb]
\includegraphics[width = \columnwidth]{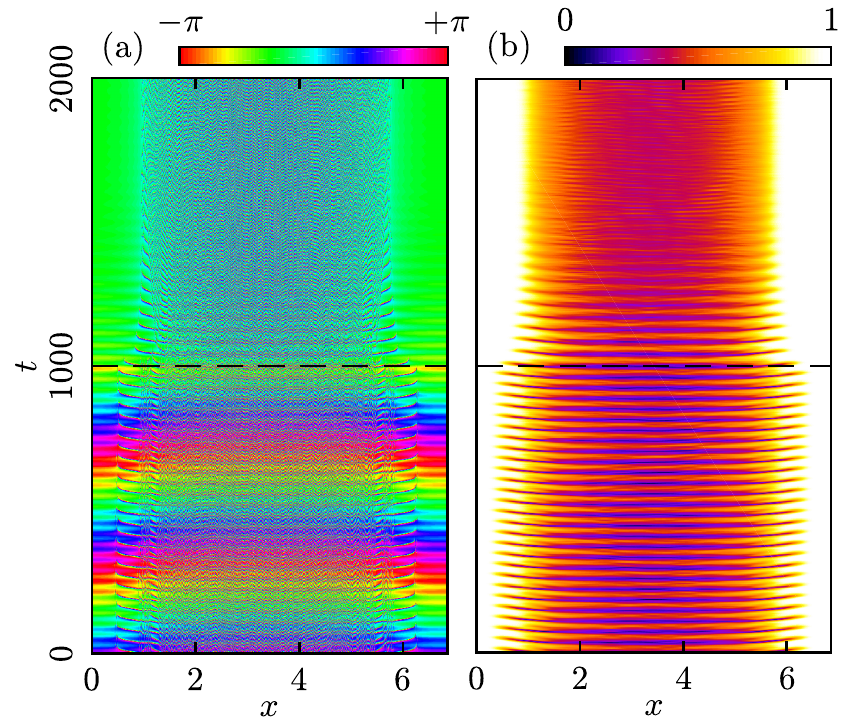}
\caption{Synchronization of breathing chimera: the spatial distribution of the 
phases $\varphi(x,t) - \Omega t$ (a), and the absolute value $\left|Z(x,t)\right|$ of 
the complex order parameter (b). Numerical 
simulations of the set of $N=8192$ oscillators were 
performed within the framework of the phase model~\eqref{eq:be} with $\varepsilon = 0.06$, $\Omega = -0.69$. 
The initial conditions was chosen in the form of a free
breathing chimera state at the length $L \approx 6.854$ of 
the oscillatory medium. The force was switched on abruptly 
at an instant of time $t_0=1000$, as marked by a black dashed straight line. 
The coarse-grained order parameter $Z(x,t)$ was calculated via local averaging with a 
Gaussian kernel $\exp\left(-x^{2}\bigl/2\varsigma^{2}\bigr.\right)$, with $\varsigma=0.1$.}
\label{fig:atbr} 
\end{figure}

\begin{figure}[ht]
\includegraphics[width = \columnwidth]{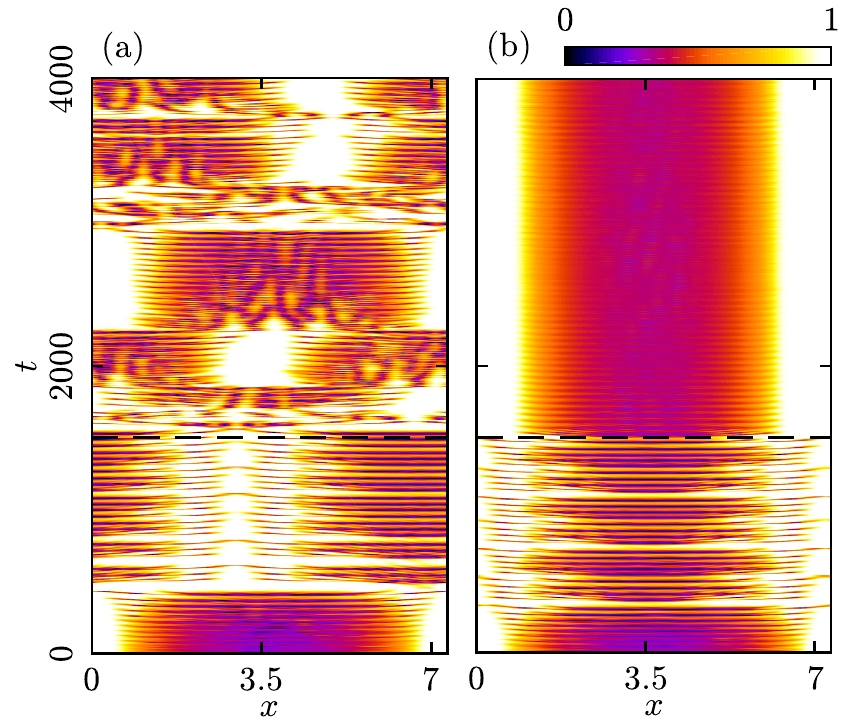}
\caption{Dynamics of amplitude of the local complex order parameter $\left|Z(x,t)\right|$ 
calculated as in Fig.~\ref{fig:atbr} according to the results of the direct numerical 
simulations of the set of $N=8192$ oscillators within the phase model~\eqref{eq:be} 
with the parameters (a) $\varepsilon = 0.03$, $\Omega = -0.68$, (b) $\varepsilon = 0.06$,  $\Omega = -0.68$.
The initial conditions was chosen in the form of a standing 
single-cluster chimera at the length $L \approx 7.332$ of 
the oscillatory medium. The force was switched on abruptly 
at an instant of time $t_0=1500$, which is marked by a  black dashed straight line.
One can see that a large enough forcing can stabilize and 
regularize the behaviour of the system which evolves to a 
standard chimera regime (see fragment (b)).}
\label{fig:atinst} 
\end{figure}

Next we consider, how the periodic force affects a breathing chimera.
The latter exists for parameters $L=6.854,\W_0=-0.69$. Here the stationary
chimera state is weakly unstable with two complex eigenvalues having
a positive real part. In the
autonomous, unforced situation, a breathing, time-periodic state appears. 
Here also the AT can be constructed as described above,
however only in a part of this locked region the constructed stationary
chimera state is stable (Fig.~\ref{fig:oscd}(a)). For very small forcing,
the locked chimera inherits the instability of the autonomous chimera, and 
evolves into a breathing state (a tiny yellow region C close to the tongue tip
in Fig.~\ref{fig:oscd}(a)). However, a large enough forcing can suppress
this ``transversal'' instability, so that in a part of the AT
(blue region A in Fig.~\ref{fig:oscd}(a))
a stationary phase-locked chimera is observed. In Fig.~\ref{fig:atbr}
this regime is shown with the phases (panel (a)) and with the order parameter (panel (b)). 
One can clearly observe the free breathing chimera up to time $t_0=1000$,
at which the forcing is switched on. Then, for $t>t_0$, both the mean frequency 
becomes locked by the force, and periodic modulations of the order parameter
disappear, what means establishing of a standard stationary chimera state.

In a part B of the AT colored red in Fig.~\ref{fig:oscd}(a), the constructed
stationary chimera state is unstable and evolves into a turbulent chimera.


Finally, we discuss regularization of a turbulent chimera. 
The latter is observed for $L=7.332$, $\W_0=-0.68$.
Here the instability of a free chimera solution is so strong that a disordered
state where the local order parameter fluctuates in space and time is observed.
The calculated AT is presented in Fig.~\ref{fig:oscd}(b).
Again, the domain of existence
of a locked stationary chimera looks like a standard triangular
synchronization domain, but only in a relatively small part (blue region)
this solution is stable. We illustrate this situation with the evolution
of the order parameter in Fig.~\ref{fig:atinst}(b). A turbulent chimera is 
observed prior to force onset time $t_0=1500$, under forcing it is 
transformed to a stable stationary chimera.

For larger values of the driving frequency (red region in Fig.~\ref{fig:oscd}(b)), 
locked solutions in presence of driving typically inherit the  instability
of the free chimera, so that
also under periodic forcing turbulent states are observed (Fig.~\ref{fig:atinst}(a)).  
We stress here, that for very large forcing
amplitudes, an observed turbulent state is a transient one, resulting
for large times in an absorbing fully synchronized regime.

Summarizing, we studied the effect of a periodic forcing on a chimera state
in a one-dimensional medium. We have constructed stationary locked
chimera patterns as periodic in space and time profiles
via solutions of a proper ordinary differential equation. The most simple
picture is observed if the free chimera is stable. Here the macroscopic
effect of forcing
on it is very similar to a general synchronization setup: there is a
locking region (AT) within which the chimera is locked by the forcing,
while outside of the AT a quasiperiodic dynamics is observed. Inside
the AT no essential microscopic and mesoscopic effects are observed.
 
Dynamics on the mesoscopic level becomes nontrivial
outside the AT, with several plateaus of locking of subgroups of oscillators
appear. On the microscopic level of the phase dynamics
of individual units, nontrivial states with rather large deviations of the phase
from the forcing one are observed despite of frequency entrainment.

Another effect not existing in simple synchronization setups is regularization
of nonstationary chimeras, breathing or turbulent. Here, inside the AT
there are subdomains, at sufficiently strong coupling,
where external forcing stabilizes stationary chimera. On the contrary, in some domains
a weakly nonstationary (breathing) chimera may become turbulent due to 
forcing. 

\acknowledgments
This paper was supported by 
the Russian Science Foundation (grant No.\ 17-12-01534)
and the Russian Foundation for Basic Research (grant No.\ 19-52-12053).
AP was partially supported by the Laboratory of Topological Methods in Dynamics NRU HSE, 
of the Ministry of Science and Higher Education of the RF, grant ag. Nr 075-15-2019-193.
The authors thank O. Omelchenko and R. G. Andrzejak for helpful discussions.

%

\end{document}